\documentclass[11pt]{article}
\usepackage{moriond,epsfig}

\def\be{\begin{equation}}
\def\ee{\end{equation}}
\def\bea{\begin{eqnarray}}
\def\eea{\end{eqnarray}}

\begin{document}
\vspace*{4cm}
\title{Prospects to measure the Higgs boson properties in ATLAS}

\author{ A. Dahlhoff }

\address{Physikalisches Institut, Universit\"at Freiburg, Germany}

\maketitle\abstracts{
As soon as a significant signal in one of the Higgs boson discovery channels is observed, it 
will be important to establish its nature. To do this, a precise measurements of its properties 
is important. In this article the prospects to measure the Higgs boson mass, width, 
spin and CP-quantum numbers, couplings to the known Standard Model particles and 
self-couplings by the ATLAS experiment are summarized.  
}

\section{Introduction}
The Standard Model (SM) is in excellent agreement with all experimental measurements. However, one of the 
important issues of the theory, the mechanism of electroweak symmetry breaking, is not 
yet verified. The Higgs mechanism is a possible solution but requires at least the experimental 
discovery of the Higgs boson. The LHC experiments, ATLAS and CMS, will be able to discover 
the SM Higgs boson in the allowed mass range from the lower limit set at LEP up to $\sim$ 1000 GeV/$\rm{c^2}$ 
for an integrated luminosity of 30 $\rm{fb}^{-1}$. After discovery, the properties of the discovered particle 
have to be investigated in detail.

\subsection{Measurement of the Higgs boson mass and width}\label{subsec:mass}
A precise measurement of the Higgs boson mass can be extracted from those channels where 
the Higgs decay products can be reconstructed, as in the case of $H \rightarrow \gamma\gamma$ and 
$H \rightarrow ZZ^{(*)} \rightarrow 4\ell$. Assuming an integrated luminosity of 300 $\rm{fb}^{-1}$ the 
Higgs boson mass can be measured directly with a precision of $\sim 0.1\%$ over the mass range 
100 - 400 GeV/$\rm{c^2}$, as shown in Fig.~\ref{fig:MassSpinCP} (left). The error is dominated by the absolute knowledge of the lepton or photon 
energy scale which is assumed to be 0.1$\%$. 
For larger masses the precision degrades due to the larger Higgs boson width and the 
increase of the statistical error. However, even for masses around 700 GeV/$\rm{c^2}$ a precision 
of about $1\%$ can be reached \cite{ATLAS-TDR}.\\
A direct measurement of the Higgs Boson width is impossible below 200 GeV/$\rm{c^2}$. For a Higgs boson mass 
above 250 GeV/$\rm{c^2}$ the detector resolution is sufficient for probing the intrinsic width with a 
precision of about $6\%$ \cite{ATLAS-TDR}.

\subsection{Measurement of the Higgs boson spin and CP eigenvalues}\label{subsec:spin}
One of the first priorities must be a determination of the spin and the CP quantum numbers. 
An analysis to probe these quantum numbers was presented in Ref.[2,3]. 
The decay mode $H \rightarrow ZZ \rightarrow 4\ell$ is used in the mass range above 200 GeV/$\rm{c^2}$ to extract 
information on spin and CP by studying two distributions of the decay correlations: the polar angle of the decay 
leptons relative to the $Z$ boson and the angle between the decay planes of the two $Z$ in the rest frame of the Higgs boson.

\begin{figure}[h!]
	  \hspace{3.0em}
          \epsfig{file=mass_600.epsi,width=0.29\linewidth}
	  \hspace{4.5em}
          \epsfig{file=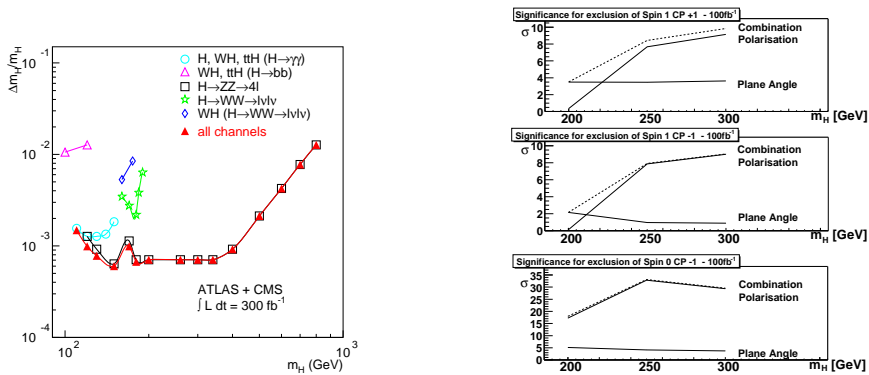,width=0.30\linewidth}
\caption{Left: The expected precision of the Higgs boson mass measurement. 
Right: Discrimination significance in $1^+$, $1^-$ and $0^-$ scenario using $H \rightarrow ZZ \rightarrow 4\ell$.
\label{fig:MassSpinCP}}
\end{figure}

In the high mass region ($\rm{m_H \ge 250 GeV/c^2}$) the angular correlations yield good discrimination power between the SM ($0^+$) Higgs boson, 
and $1^+$, $1^-$, $0^-$ scenarios (see Fig.~\ref{fig:MassSpinCP} right). In this context it should be noted that the 
Spin 1 hypothesis is also ruled out by observing non-zero $H\gamma\gamma$ and $Hgg$ couplings.    
All scenarios considered can be separated from the SM with a significance of more than 8 $\sigma$ 
with an integrated luminosity of 100 $\rm{fb}^{-1}$. The decay plane angle correlation becomes more important for 
lower Higgs boson masses, where the discrimination power of the polar angle variable decreases. 
For a Higgs boson mass below 200 GeV/$\rm{c^2}$ information on spin and CP can be extracted from the azimuthal separation 
of the leptons in the vector boson fusion process $qq \rightarrow qqH \rightarrow qqWW \rightarrow qq \ell\nu \ell\nu$ \cite{SAsai}.

\subsection{Measurement of the Higgs boson coupling parameters}\label{subsec:coupl}
Based on all ATLAS simulation studies for a SM Higgs boson, a global Maximum Likelihood fit has beem performed 
to determine the expected accuracy on the measurement of the coupling parameters. 
So far, the interesting low mass region 110 GeV/$\rm{c^2}$ to 190 GeV/$\rm{c^2}$ has been considered in Ref.\cite{MDuehr1}.

\begin{table}[h!]
   \begin{center}
   \footnotesize
   \begin{tabular}{@{}l@{$\,$}l|l@{\hspace{0.3ex}}|c}
      \multicolumn{2}{c|}{Production}&\multicolumn{1}{c|}{Decay}&Mass range\\ \hline
      &Gluon Fusion &$H \rightarrow ZZ^{(*)} \rightarrow 4\ell$    &110 GeV - 200 GeV\\
      &($gg \rightarrow H$) &$H \rightarrow WW^{(*)} \rightarrow \ell \nu \, \ell \nu$    &110 GeV - 200 GeV\\
      & &$H \rightarrow \gamma \gamma$    &110 GeV - 150 GeV\\ \hline
      &Weak Boson & $H \rightarrow ZZ^{(*)} \rightarrow 4\ell$    &110 GeV - 200 GeV\\
      &Fusion &$H \rightarrow WW^{(*)} \rightarrow \ell \nu \, \ell \nu$    &110 GeV - 190 GeV\\
      &     &$H \rightarrow \tau \tau \rightarrow \ell \nu \nu \, \ell \nu \nu$ ($\ell \nu \nu \, \rm{had} \nu$)   &110 GeV - 150 GeV\\
      &           &$H \rightarrow \gamma\gamma$    &110 GeV - 150 GeV\\ \hline
      &$t\bar{t}H$   &$H \rightarrow WW^{(*)} \rightarrow \ell \nu \, \ell \nu\, (\ell \nu)$    &120 GeV - 200 GeV\\
      &              &$H \rightarrow b\bar{b}$    &110 GeV - 140 GeV\\
      &              &$H \rightarrow \gamma \gamma$    &110 GeV - 120 GeV\\ \hline
      &$WH$          &$H \rightarrow WW^{(*)} \rightarrow \ell \nu \, \ell \nu\, (\ell \nu)$    &150 GeV - 190 GeV\\
      &              &$H \rightarrow \gamma \gamma$    &110 GeV - 120 GeV\\ \cline{2-4}
      &$ZH$          &$H \rightarrow \gamma \gamma$    &110 GeV - 120 GeV\\ 
   
   \end{tabular}
   \end{center}
\caption{List of all ATLAS studies used for the Maximum Likelihood fit. 
The mass range is the range of Higgs boson mass considered in the studies (not the discovery region).
\label{TabStudies}}   
\end{table}
\normalsize

\begin{figure}[h!]
          \epsfig{file=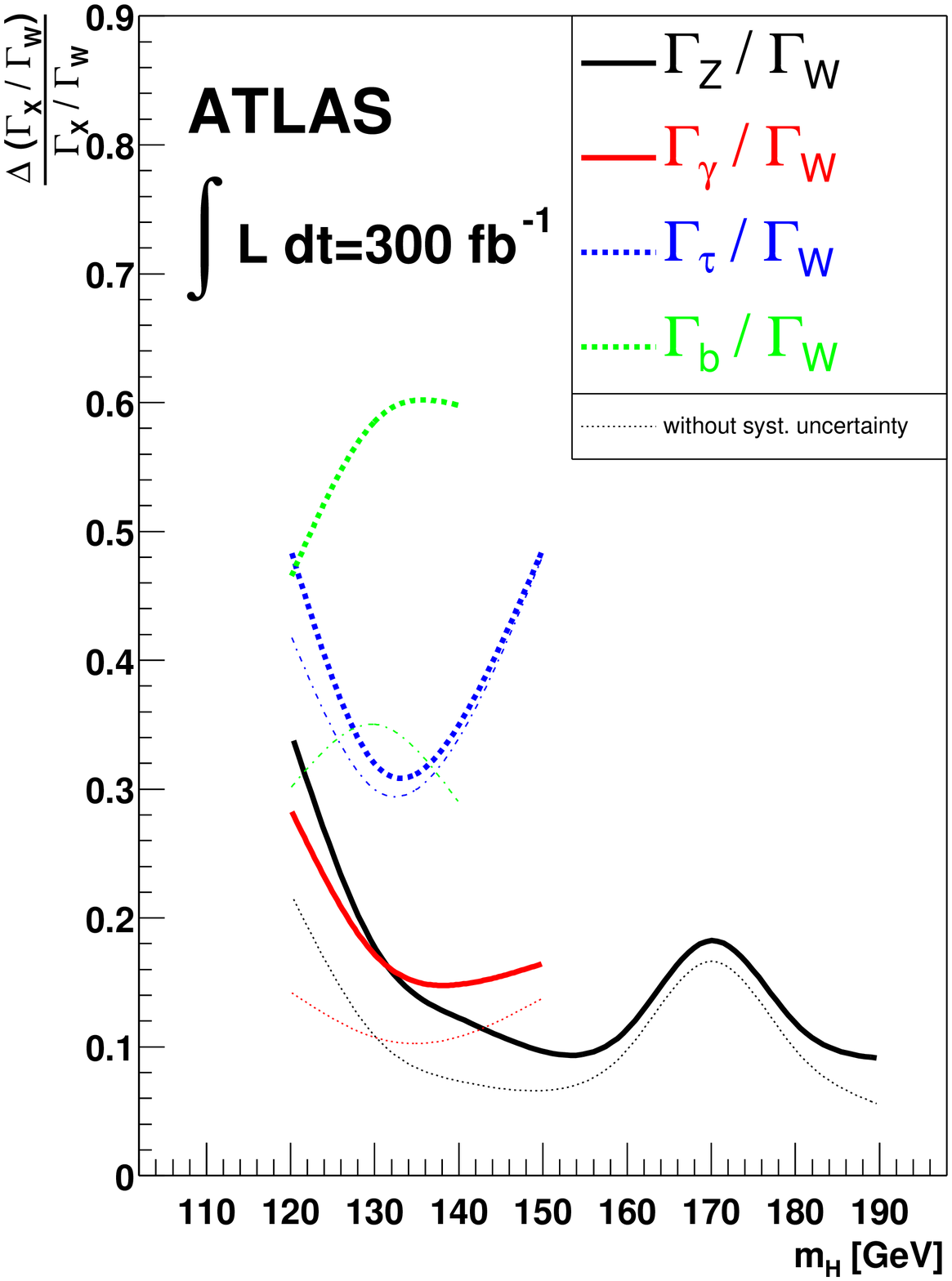,width=0.3\linewidth}
          \epsfig{file=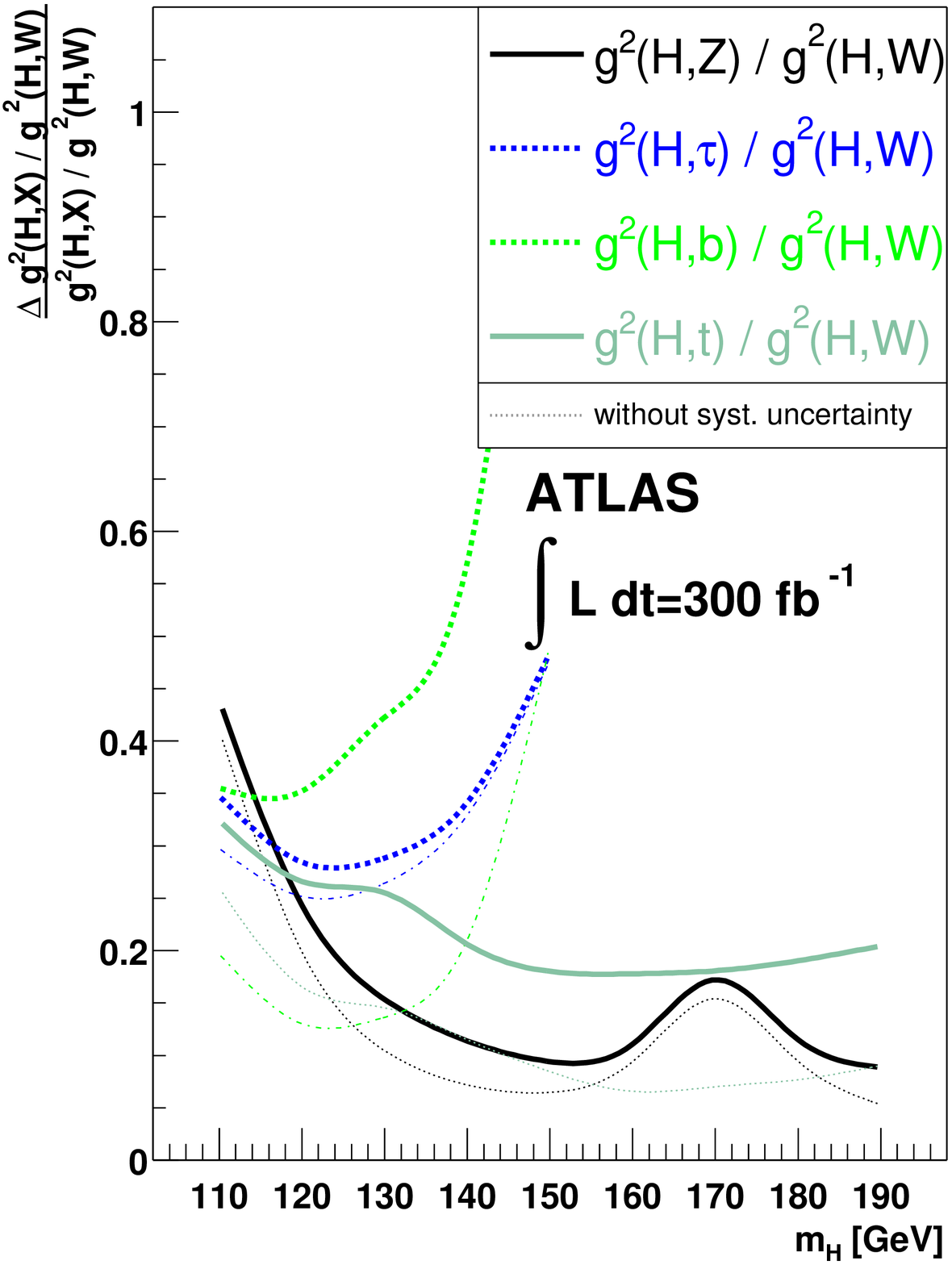,width=0.3\linewidth}
          \epsfig{file=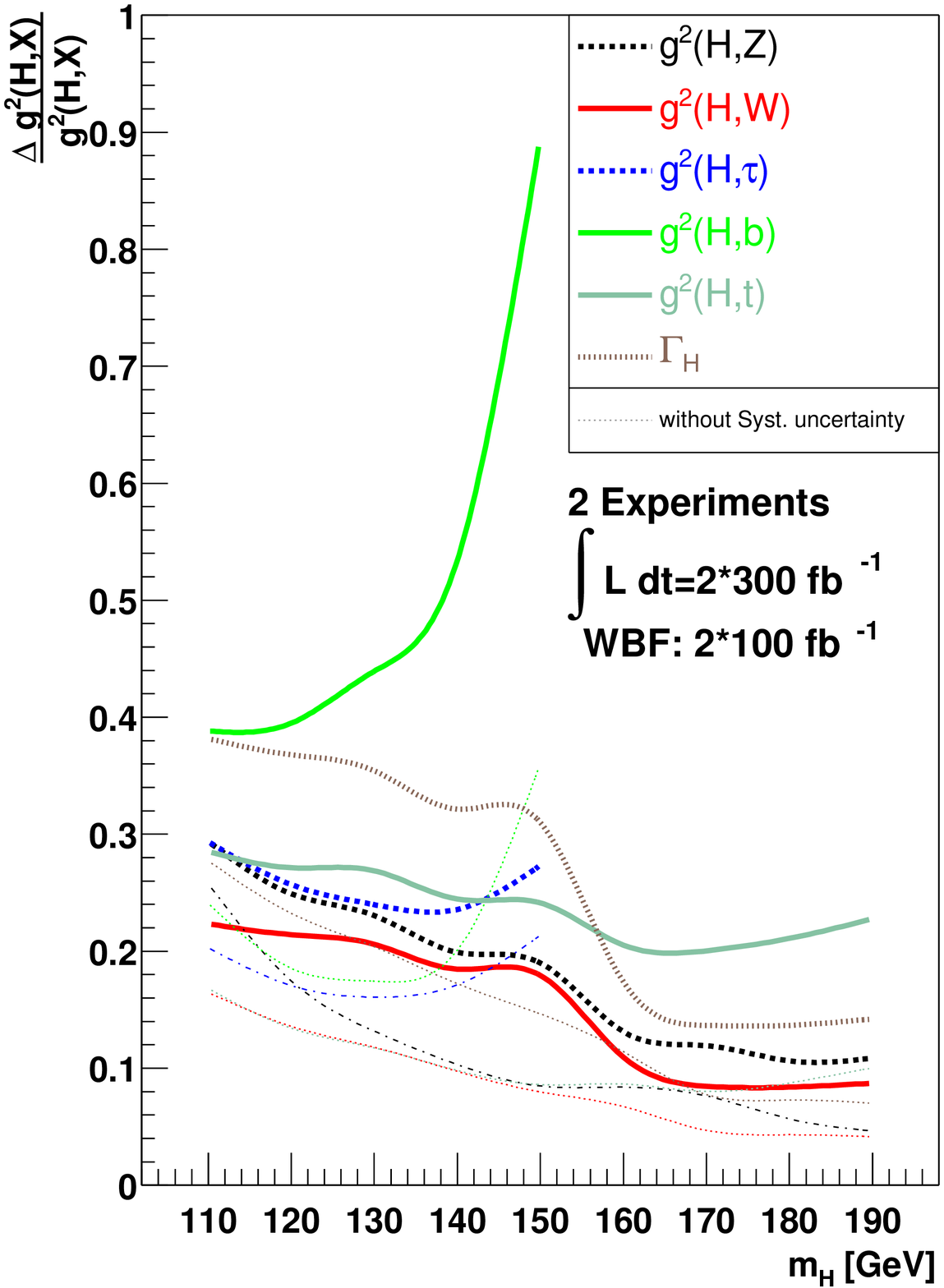,width=0.296\linewidth}
\caption{Expected relative error on the measurement of ratios of partial widths, coupling ratios and absolute couplings. 
The dotted lines give the expectation without systematic uncertainties.
\label{fig:CouprelBR}}
\end{figure}

The fit takes experimental and theoretical systematic errors into 
account. Depending on theoretical assumptions different kinds of coupling parameters can be extracted. 
All fits are based on a CP-even, spin-0 Higgs boson. Hypothesis assuming that only one light 
Higgs boson exists, it is possible to measure ratios of Higgs boson partial widths. 
The assumption that there are no new particles involved in $gg \rightarrow H$ production 
and no extremely enhanced couplings to light fermions allows to measure ratios of Higgs Boson 
couplings of $Z$ boson, $\tau$, $b$ and $t$ fermion to the $W$ boson coupling.
For an integrated luminosity of 300 $\rm{fb}^{-1}$, the ratios can be measured to $10\%$ - $30\%$.
With the assumption of an upper limit on the $W$ and $Z$ coupling and of a lower limit on $\Gamma_H$, 
an absolute measurement of coupling parameters is possible \cite{MDuehr2}. The expected relative accuracy (Fig.~\ref{fig:CouprelBR}) 
for this measurement is between $10\%$ and $40\%$. 

\subsection{Probing the Higgs boson self-coulings at LHC (SLHC)}\label{sec:selfcoupl}
\begin{figure}[h!]
\begin{center}
          \epsfig{file=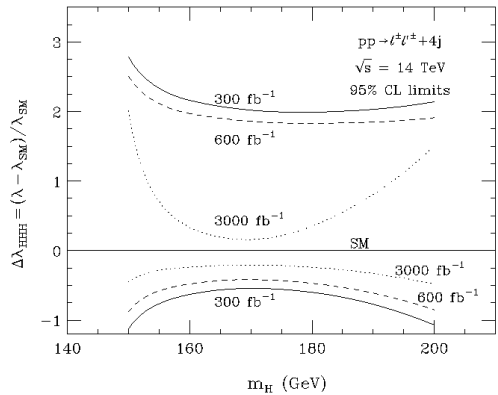,width=0.45\linewidth}\vspace{-0.5em}
\end{center}	  
\caption{Sensitivity bounds for the Higgs boson self-coupling are shown for several integrated luminosity scenarios (taken from Ref.[7]). 
\label{fig:SelfCoup}}
\end{figure}

The most important task after a Higgs boson discovery and coupling measurements is the extraction of the Higgs potential. This requires the  measurement 
of the trilinear and quartic Higgs Boson self-coupling. Only multiple Higgs boson production can probe these directly.
Since the quartic coupling is about two orders of magnitude smaller than the trilinear coupling, present studies 
have focussed on the determination of $\lambda_{HHH}$. Studies have shown that the 
$HH \rightarrow WW WW \rightarrow (\ell^{\pm}\nu jj) (\ell^{\pm}\nu jj)$ final state has the highest sensitivity for extracting 
information on the self-coupling parameter. 
A recent study \cite{Baur}, based parton level calculations, concludes that the LHC may rule out the case 
of a non-vanishing $\lambda_{HHH}$ within a mass range of 150 $\rm{GeV/c^2}$ $\le$ $m_H$ $\le$ 200 $\rm{GeV/c^2}$ with a confidence 
level of 95$\%$. For the upgraded LHC, assuming an integrated luminosity of 3000 $\rm{fb}^{-1}$ per experiment, it is claimed that 
a measurement of $\lambda_{HHH}$ with 20$\%$ precision will be possible (Fig.~\ref{fig:SelfCoup}). In order to derive more realistic sensitivity bounds for 
the Higgs boson self-coupling, more detailed simulations, which take the detector performance into account, are needed. 
First preliminary studies were performed to confirm the sensitivity at the SLHC scenario \cite{FMazz}. 
However, some background contributions might have been underestimated. Further studies on this subject are currently in progress \cite{ADahl}. 

\section{Conclusion}
The expected performance of both LHC experiments offers a good possibility not only to discover a Higgs boson but 
also to measure its properties like mass, spin/CP and couplings. For typical mass regions the precision 
of the mass measurement is about 0.1$\%$. For a Higgs boson mass above 200 GeV/$\rm{c^2}$ spin/CP sensitive measurements are 
possible using angular correlations in the $gg \rightarrow H \rightarrow ZZ$ channel. For the critical region of Higgs boson masses 
below 200 GeV/$\rm{c^2}$ ratios of couplings squared and absolute couplings squared can be 
extracted with 10$\%$ to 50$\%$ relative error using moderate theoretical assumptions. A first probe of the Higgs
 self-coupling to determine the shape of the Higgs potential might be possible at a luminosity upgraded LHC.

\end{document}